\documentclass[aps,prl,twocolumn,showpacs,longbibliography]{revtex4-1}
\usepackage{amsmath}
\usepackage{amssymb}
\usepackage{graphicx}
\usepackage{color}

\definecolor{forestgreen}{rgb}{0.13,0.54,0.13}
\definecolor{grey}{rgb}{0.6,0.6,0.6}
\newcommand{\comments}[1]{}

\newcommand{\rmc}{{\rm c}}
\newcommand{\rme}{{\rm e}}
\newcommand{\rmd}{{\rm d}}
\newcommand{\rmi}{{\rm i}}

\newcommand{\rmq}{{\rm q}}

\begin{document}

\title{Permanent spin currents in cavity-qubit systems}

\author{Manas Kulkarni}
\affiliation{International Centre for Theoretical Sciences, Tata Institute of Fundamental Research, Bangalore - 560012, India}
\author{Sven M. Hein}
\affiliation{Technische Universität Berlin, Institut f\"ur Theoretische Physik, Nichtlineare Optik und Quantenelektronik, Hardenbergstraße 36, 10623 Berlin, Germany}
\author{Eliot Kapit}
\affiliation{Department of Physics and Engineering Physics, Tulane University, New Orleans, LA 70118, USA}
\author{Camille Aron}
\affiliation{Laboratoire de Physique Th\'eorique, \'Ecole Normale Sup\'erieure, CNRS, \\
PSL Research University, Sorbonne Universit\'es, 75005 Paris, France}
\affiliation{Instituut voor Theoretische Fysica, KU Leuven, Belgium}

\begin{abstract}
In a recent remarkable experiment [P. Roushan \textit{et al.}, Nature Physics \textbf{13}, 146 (2017)], a spin current in an architecture of  three superconducting qubits  was produced during a few microseconds by creating synthetic magnetic fields.  The life-time of the current was set by the typical dissipative mechanisms that occur in those systems. We propose a scheme for the generation of \emph{permanent} currents, even in the presence of such imperfections, and scalable to larger system sizes. It relies on striking a subtle balance between multiple nonequilibrium drives and the dissipation mechanisms, in order to engineer and stimulate chiral excited states which can carry current.

\end{abstract}

\maketitle

\textit{Introduction:}
Understanding and engineering transport properties of mesoscopic and condensed-matter systems is useful both from a fundamental perspective (\textit{e.g.} understanding  quantum impurity physics \cite{takis_kondo,schiro1,het1}) and from the point of view of device applications (\textit{e.g.} quantum diodes, rectifiers and transistors \cite{phonon0,JIANG20161047}). More recently transport in quantum information technologies has been of great interest as uninterrupted, fast, and reliable transmission of information is key to quantum computation and simulation schemes~\cite{nori,natphys0,natphys1}. 

Recenly, there has been remarkable progress in engineering quantum dot circuit-QED systems where electronic transport is harnessed as a gain medium~\cite{Liu285, mk0,mk1,mk2} and helps in the realization of novel quantum devices such as microwave amplifiers and lasers in microwave regime. Similarly, the role of phonons in making thermoelectric transistors and rectifiers in quantum-dot based systems is actively investigated~\cite{phonon0}. Quantum dot cQED architectures, however, are difficult to scale up to many qubit systems. Similar advances have also been made in transmon-resonator cQED architectures; results of particular relevant to this work include the passive stabilization of arbitrary single qubit states~\cite{murchvool2012,hollandvlastakis2015,hacohenmartin2016}, and more complex many-qubit entangled states~\cite{leghtasvool2013,shankarhatridge2013,gourgyramasesh2015,schwartzmartin2016,maowens2017}, using engineered dissipation.

On the other hand, creating controlled photonic currents is even more challenging because of the lack of a chemical potential for photons. One could, in principle, subject a photonic system to a finite temperature bias to generate a photonic Seebeck current, but this remains a fantastic experimental ambition~\cite{ap0,ap1}.
Among the recent progresses in the engineering and the control of Hybrid Quantum Systems, the creation of synthetic magnetic fields for photons opens the door to investigating quantum hall physics~\cite{qh0,qh1,kapithafezi2014,andersonma2016}, and other exotic quantum many-body phenomena~\cite{km0} in mesoscopic setups. In a recent remarkable experiment~\cite{roushan} proposed by one of the authors~\cite{kapit_pra}, a synthetic magnetic field was successfully produced in a superconducting-qubit architecture, generating a current in a three-qubit ring (of a magnitude of $10^{6}$ excitations per seconds). However, due to  unavoidable environmental dissipative mechanisms, the current could only be observed for a few microseconds, thereby resulting in only a few excitations transported during the experiment.

From the point of view quantum computation and quantum simulation, passive generation and stabilization of many-qubit states has long been an important goal~\cite{kapit2017review}. In these cases, natural dissipative processes, which become more and more detrimental to quantum coherence as the system complexity is increased, are balanced by adding cleverly tuned dissipative sources, which are capable of passively correcting unwanted processes from intrinsic dissipation, and generating states from vacuum. In this Letter, we leverage these techniques to show how to achieve indefinitely long-lived currents in a ring of superconducting qubits. We use a delicate interplay of drive sources and the unavoidable dissipative mechanisms to stabilize a current-carrying chiral non-equilibrium steady state.

The combined use of driving and dissipation to realize non-trivial quantum states has already been successfully implemented  in superconducting-qubit architectures (see Ref.~\cite{kapit2017review}, and references therein). Recently, some of the authors proposed a scheme to generate high-fidelity quantum entanglement between distant qubits~\cite{dimerpra,cami2}. The concept and theoretical framework were later validated experimentally, with a singlet state sustained indefinitely~\cite{mollie}. It was also shown that this scheme is scalable~\cite{camiprx}, a promising feature given the ongoing experimental efforts to increase the system sizes in this field~\cite{natphys}. 

Here, we adapt those recent ideas and theoretical methods to a different context: the production of a synthetic magnetic field for photons and the generation of a  photon-assisted permanent current of qubit excitations. We obtain currents on the order of the mega-excitations per second. Our results also demonstrate that persistent currents and long-lived entangled states are deeply connected, and any progress on one side will benefit the other side.

\begin{figure}
\begin{center}
 \includegraphics[width=0.7\hsize]{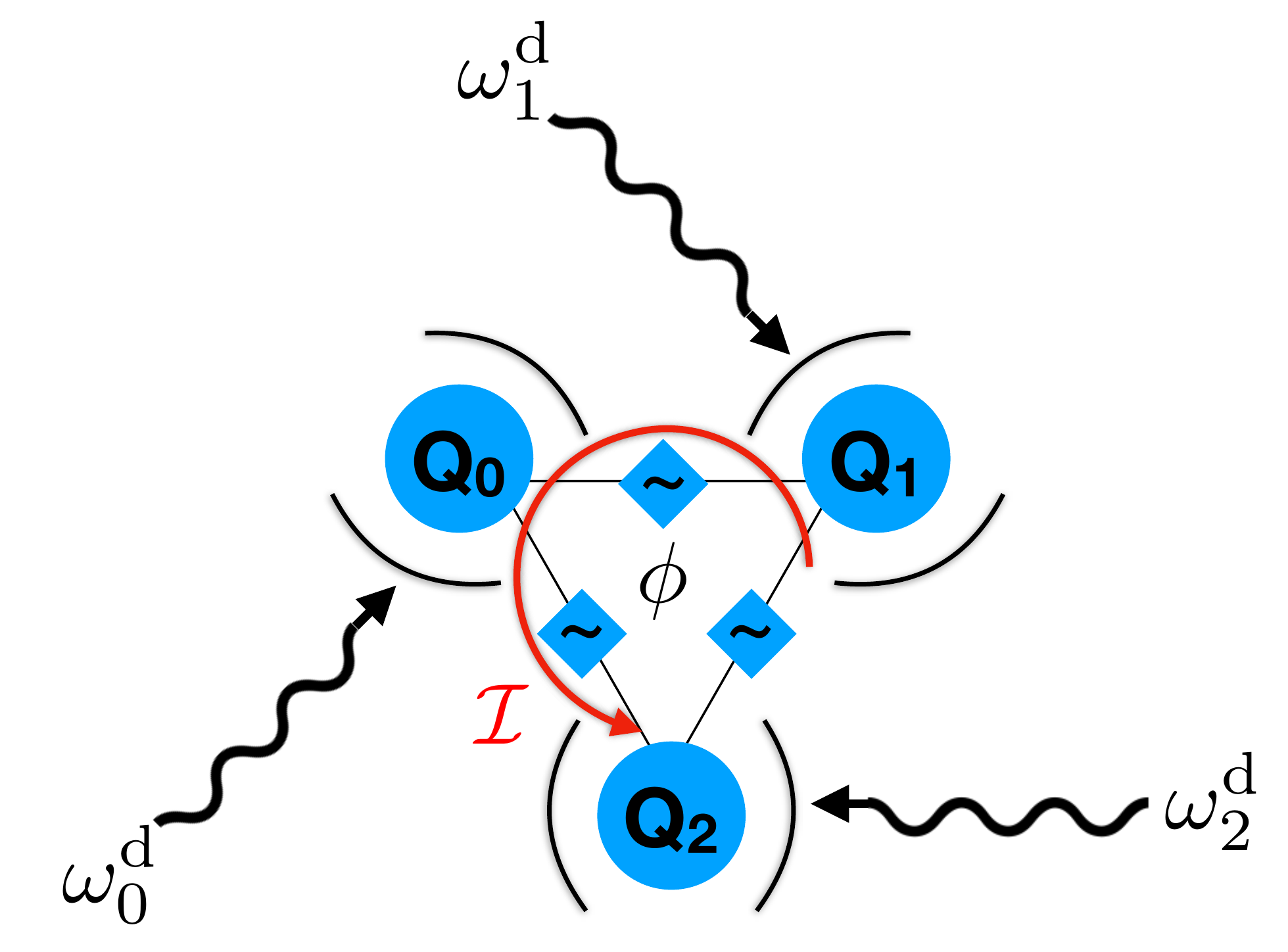}
 \caption{N=3 qubits, $Q_0$, $Q_1$, and $Q_2$, are coupled together in a ring geometry via time-dependent couplers whose phase $\phi$ creates an artificial magnetic flux. The latter is the main force to drive a current $\mathcal{I}$ of qubit excitations. To balance the dissipative losses, and generate a persistent current, the qubits are also coupled to optical cavities driven by microwave sources with carefully selected frequencies $\omega^\rmd_i$, $i = 0, 1, 2$.
}
 \label{fig:setup}
\end{center}
\end{figure}

\textit{Model}: We consider an open quantum system variant of a recent experimental setup~\cite{roushan}. As shown in Fig.~\ref{fig:setup}, it consists of $N$ qubits (artificial two-level systems) arranged in a ring geometry, and capacitively connected by time-dependent couplers. Each qubit  is also coupled to a photonic cavity which is coherently driven by an independent microwave source. Note that it is also possible to have only one of the qubits coupled to a driven cavity.
In this Letter, we present analytical results for a generic ring of size $N$. However, in order to make a strong connection with the recent experiment of Ref.~\cite{roushan}, we display figures in the particular case of $N=3$. The time-dependent Hamiltonian is given by (we set $\hbar = 1$)
\begin{align} \label{eq:H}
 H(t) = H_\sigma(t) + H_a(t) + H_{\sigma a} \,,
 \end{align}
where $H_\sigma(t)$, $H_a(t)$, and $H_{\sigma a}$ are respectively the coupled qubits, the driven cavities, and the cavity-qubit couplings Hamiltonians:
 \begin{align}
H_\sigma(t) =& \! \sum_i   \omega^\rmq_i \frac{\sigma_i^z}{2} - J_i(t)  \left[ \sigma_i^+ \sigma_{i+1}^- +  \mbox{h.c.} \right],\,  \\
H_a(t) =& \! \sum_i  \omega^\rmc_i a_i^\dagger a_i 
  +  2 \epsilon^{\rmd}_i \cos(\omega^\rmd_i t + \Phi_i^\rmd) \left[ a_i + a^\dagger_i \right],
\\ H_{\sigma a} =&   \! \sum_i  g \,  \sigma^x_i \left[ a_i + a_i^\dagger \right]
\,. \label{eq:Ha}
\end{align}
Importantly, the qubits have different energy splittings $\omega^\rmq_i$ at different sites. We write $\omega^\rmq_i = \omega_\rmq + \delta_i$.
Their time-dependent nearest-neighbor coupling is given by $J_i(t)\ = 2 J_0 \cos(  \Delta_i t + \phi_i)$ where the amplitude $J_0 \in \mathbb{R}$ and the driving frequency $\Delta_i$ will be set below.
For simplicity, the phases of the couplers are taken to be equal for all bonds, $\phi_i = \phi \in [0,2\pi )$. %*** Explain:
We take the cavities to be fabricated such that their mode frequencies are set by $ \omega^{\rmc}_i = \omega_{\rmc} +  \delta_i$, thus ensuring the cavity-qubit detuning $\omega_i^\rmq - \omega_i^\rmc \equiv \Delta$ to be site independent.  $\epsilon^\rmd_i$, $ \omega^\rmd_i$, $\Phi_i^\rmd$ are respectively the amplitude, frequency, and phase of each cavity microwave drive. 
Our computations show that the site-dependence of the driving amplitudes and phases does not play any crucial role in what follows.
Therefore, for simplicity, we present our results when all cavities are driven with the same amplitude $\epsilon_i^\rmd = \epsilon_\rmd$, and the same phase $\Phi_i^\rmd = 0$.
The light-matter coupling, $H_{\sigma a}$, is of the Rabi type and its strength is controlled by the dimensionless ratio $g/\Delta$.

Qubits and cavities are also subject to inevitable spontaneous dissipative mechanisms:  decay of qubit excitations, intrinsic qubit dephasing, and cavity damping, occurring at rates $\gamma$, $\gamma_\phi$, and $\kappa$, respectively. 
As we shall see later, the qubit degeneracies are lifted on a scale set by the qubit-qubit coupling, implying that the phase noise spectrum (typically $1/f$-like) is probed at a finite frequency $J_0$ rather than DC. This  strongly suppresses  $\gamma_\phi$ compared to uncoupled qubits (see Refs.~\cite{mollie,roushan,kapit2017review} for details).
%We use a white noise model for dephasing, appropriate to the situation at hand, where a continuously many-qubit Hamiltonian is applied which breaks time reversal symmetry. This suppresses local low frequency ($1/f$-like) phase noise, reducing in an empirical model well-approximated by simple white noise phase errors with a reduced rate (see Refs.~\cite{roushan,kapit2017review} for details).

In modern superconducting architecture, a good phenomenological set of parameters is: $\omega_\rmq \approx 7$,  $\omega_\rmc \approx 6$,  $g = 10^{-1}$, $|\delta_i| \approx 10^{-2}$, $J_0 \approx 10^{-3}$, $\kappa \approx 10^{-4}$, $\gamma \approx 10^{-5}$, and $\gamma_\phi \approx 10^{-6}$, all in units of $2\pi~$GHz. These are the values that we use in our computations. They correspond to the hierarchy $ \Delta \gg g \gg |\delta_i| \gg J_0 \gg \kappa \gg \gamma \gg  \gamma_\phi$ that guides the different approximations in our analytic framework.

By setting the driving frequency of the qubit couplers to match the energy difference between adjacent qubits, \textit{i.e.} $\Delta_i = \delta_i - \delta_{i+1} \neq 0$, together with driving the cavities at the frequencies $\omega^\rmd_i = \omega_\rmd + \delta_i$, the Hamiltonian~(\ref{eq:H}) can be rendered time-independent by moving to a rotating frame \textit{via} the $\mathfrak{U}(1)$ transformation $U(t) \equiv \prod_{i} \exp\left[ \rmi \omega^\rmd_i t \left( {\sigma^z_i}/{2} + a^\dagger_i a_i \right) \right] $, and neglecting the high-frequency counter-rotating terms.

The light-matter interaction, $H_{\sigma a}$, is treated with a standard Schrieffer-Wolff transformation, \textit{i.e.} a second-order perturbation in $g/\Delta$. Altogether, we obtain
 \begin{align}
H_\sigma =& \! \sum_i  
\boldsymbol{h_i} \cdot \frac{\boldsymbol{\sigma}_i}{2} -J_0 \left[
 \rme^{\rmi \phi}\sigma^+_{i}  \sigma^-_{i+1} + \mbox{h.c.}
 \right]\,, \label{eq:Hspinchain} \\
H_{\sigma a} = & \left(\frac{g}{\Delta}\right)^2  \sum_i  \frac{\sigma_i^z}{2} \left[ \Delta a_i^\dagger a_i + \epsilon_\rmd a_i^\dagger   + \mathrm{h.c.} \right] \,,  \\
H_a =&  \sum_i \big[ - \Delta_\rmc a^\dagger_i a_i  +  \epsilon_\rmd  (a_i + a^\dagger_i  ) \big] \,,
\label{eqtransXY}
\end{align}
with ${h}^x_i = 2  (g/\Delta) \epsilon_\rmd $, $h^y_i= 0$, $h^z_i  = \Delta_\rmq + \delta\omega_\rmq  $ where $\Delta_\rmq \equiv \omega_\rmq - \omega_\rmd$, $\Delta_\rmc \equiv \omega_\rmd - \omega_\rmc$, and  $\delta\omega_\rmq =  \Delta (g/\Delta)^2$  is the cavity-induced Lamb shift. 
We neglected the emergent qubit-mediated cavity-cavity interactions, which are virtual processes occurring at energy scales much smaller than the cavity linewidth, namely $ J_0  \left({g}/{\Delta}\right)^2 \ll \kappa$. 

Note that we started with a system that is not translationally invariant (qubits and cavities are different at each site) nonetheless, once driven appropriately,
 the effective description of the model is now translationally invariant with two main independent drive parameters,  $\omega_\rmd$ and $\phi$, that can be easily tuned \textit{in situ}.

In order to diagonalize the qubit sector of the Hamiltonian, $H_\sigma$, we first simplify the problem by truncating its Hilbert space to the zero-energy ground state  $|\boldsymbol{0} \rangle  \equiv | \downarrow \ldots \downarrow \rangle$ and the states of the single-excitation manifold, $|i\rangle \equiv |\downarrow_0 \ldots\downarrow_{i-1} \, \uparrow_i \, \downarrow_{i+1} \ldots \downarrow_{N-1}  \rangle$. The truncated $H_\sigma$ reads
\begin{align} \label{eq:Hstrunc}
H_\sigma  = \!\! \sum_k \! E_k |k \rangle \langle k | + \left( \frac{g}{\Delta} \right) \sqrt{N} \epsilon_\rmd   \left(  | k=0 \rangle\langle \boldsymbol{0} | + \mbox{h.c.} \right),
\end{align} 
where $|k\rangle \equiv 1/\sqrt{N}  \sum_{i=0}^{N-1}  \rme^{\rmi k i} \ |i\rangle$ are the chiral one-excitation eigenstates of $H_\sigma$ when turning off the cavity drives ($\epsilon_\rmd = 0$). They carry a single qubit excitation of quasi-momentum $k= 2\pi\, n/N$, with $n = 0\ldots N-1$, delocalized over the entire ring, with a dispersion relation
$E_k = \epsilon_k - \omega_\rmd, \; \epsilon_k = \omega_\rmq + \delta\omega_\rmq - 2 J_0  \cos(k + \phi)$.
Here, the Lamb shift was renormalized by the cavity drives: $\delta\omega_\rmq \simeq  (g/\Delta)^2 \Delta \left[ 1 + 12  (\epsilon_\rmd/\Delta)^2\right]$.

We then diagonalize the qubit sector by perturbation theory in the lowest order in $({g}/{\Delta})( \epsilon_\rmd/\Delta_\rmq)$. The spectrum of $H_\sigma$ reads
\begin{align}
|\widetilde{\boldsymbol{0}} \rangle 
 &\simeq
  |\boldsymbol{0} \rangle \!-\! \left(\frac{g}{\Delta}\right)  \! \frac{\sqrt{N} \epsilon_\rmd}{\Delta_\rmq} |k =0 \rangle,\, 
\widetilde{E}_{\boldsymbol{0}} \! \simeq \!-\!  \left(\frac{g}{\Delta}\right)^2  \! \frac{N \epsilon_\rmd^2}{\Delta_\rmq}, \label{eqqbitcollen} \\
|\widetilde{k} \rangle & \simeq
|k \rangle \! + \! \delta_{k,0} \left(\frac{g}{\Delta}\right) \! \frac{\sqrt{N} {\epsilon_\rmd}}{\Delta_\rmq} |\boldsymbol{0} \rangle,\,
\widetilde{E}_k \! \simeq \! E_k \! - \! \frac12 \! \left(\frac{g}{\Delta}\right)^2  \! \frac{N \epsilon_\rmd^2}{\Delta_\rmq}. \nonumber
\end{align}
For the consistency of the perturbation theory,  we also included the corrections of order $(g/\Delta)^2$ to $\widetilde{E}_k$ that are due to the coupling to double-excited states. They may be estimated by a standard two-magnon coordinate Bethe ansatz computation.
Since we are mostly concerned with the dynamics of the qubits, we integrate over the cavity degrees of freedom by linearizing the photon excitations around a classical background, $a_i \equiv \bar a + d_i$ with  $\bar a = \frac{\epsilon_\rmd}{\omega_\rmd - \omega_\rmc + \rmi \kappa/2}$, and we later treat the dynamics induced by the quantum fluctuations, \textit{i.e.} the $d_i$'s, \textit{via} a Fermi Golden Rule approach.
Neglecting those light-matter interaction terms that are quadratic in the fluctuations, the linearized light and light-matter sectors reduce to 
\begin{align}
H_a \rightarrow H_d &= \sum_i (\omega_\rmc - \omega_\rmd) d_i^\dagger d_i \,, \label{eq:bath1} \\
H_{\sigma a} \rightarrow H_{\sigma d} &= \left(\frac{g}{\Delta}\right)^2  \sum_i  \sigma^z_i ( \Delta \bar{a}  +\frac{1}{2} \epsilon_\rmd) d_i^\dagger + \mbox{h.c.} \,. \label{eq:bath2}
\end{align}
Assuming that the photon fluctuations are thermalized close to zero temperature, we integrate them out by employing the Fermi Golden Rule. The Born approximation is valid due to the small residual light-matter coupling in the Hamiltonian $H_{\sigma d}$. The Markov approximation becomes exact in the steady state since the typical timescale of variation of the reduced density matrix for the qubits,  $\rho_\sigma(t) \equiv \mbox{Tr}_d \left[ \rho(t) \right] $, is always much larger than the timescale of the photon fluctuations, $1/\kappa$. 
In the steady state,  $\rho_\sigma^{\infty} \equiv \lim\limits_{t\to\infty} \rho_\sigma(t)$, the driven-dissipative dynamics are given by the following Master Equation
\begin{align}\label{eq:master}
\partial_t \rho^{\infty}_\sigma & \! \! =  0 =  -\rmi \left[ H_\sigma,\rho^{\infty}_\sigma \right]
+ 
\sum_k \Gamma_{\boldsymbol{0}\to k} \mathcal{D}[| \widetilde{k} \rangle \langle \widetilde{\boldsymbol{0}} | ] \rho^{\infty}_\sigma \nonumber \\
& \hspace{-0em} + \gamma \sum_k  \mathcal{D}[| \widetilde{\boldsymbol{0}} \rangle \langle \widetilde{k} | ]  \rho^{\infty}_\sigma
 +  \frac{2\gamma_\phi}{N} \sum_{k\,q} \mathcal{D}[| \widetilde{q} \rangle \langle \widetilde{k} | ] \rho^{\infty}_\sigma.
\end{align}

The qubit decay terms $\gamma \, \mathcal{D}[| \widetilde{\boldsymbol{0}} \rangle \langle \widetilde{k} | ]$, with the Lindblad operator $\mathcal{D}[X]\rho \equiv (X \rho X^\dagger - X^\dagger X \rho + \mathrm{h.c.})/2 $, describe the spontaneous relaxation from single-excited states to the ground state. The qubit dephasing terms describe all-to-all transitions between states in the single-excitation manifold.
The pumping rates induced by the photon fluctuations are
\begin{align}\label{eq:rate}
\Gamma_{\boldsymbol{0} \to k}= 2 \pi    \Lambda^2 \, \rho(\omega_\rmd+\widetilde{E}_{\boldsymbol{0}} -\widetilde{E}_k)\,,
\end{align}
where the cavity density of states is the Lorentzian
$\rho(\omega) = - \pi^{-1} \mbox{Im } [ {\omega -\omega_\rmc +\rmi\kappa/2} ]^{-1 }$,
and the transition matrix element is given by 
$\Lambda^2 =    ({g}/{\Delta})^6
 (1+2 {\Delta}/{\Delta_\rmc})^2
 {\epsilon_\rmd^4}/{\Delta_\rmq^2} $.
To maximize the transition rate $\Gamma_{\boldsymbol{0} \to k}$, $\omega_\rmd$ should satisfy the energy conservation
$\omega_\rmd+\widetilde{E}_{\boldsymbol{0}} -\widetilde{E}_k = \omega_\rmc$,
\textit{i.e.} the optimal $\omega_\rmd$ is set by
\begin{align} \label{eq:wopt}
  \omega_{\rmd}^{\rm opt}(\phi)  = \bar\omega_\rmd  - J_0 \cos(k+\phi) \,,
\end{align}
with $2 \bar\omega_\rmd \approx \omega_\rmc + \omega_\rmq + \delta \omega_\rmq +  \left({g}/{\Delta}\right)^2   N \epsilon_\rmd^2 / \Delta$. This expression reveals the Raman inelastic scattering nature of the process: two incoming photons contribute to creating a qubit excitation and dumping the excess energy in a cavity photon.

\begin{figure}
\begin{center}

 \includegraphics[height=0.41\hsize, angle=-0]{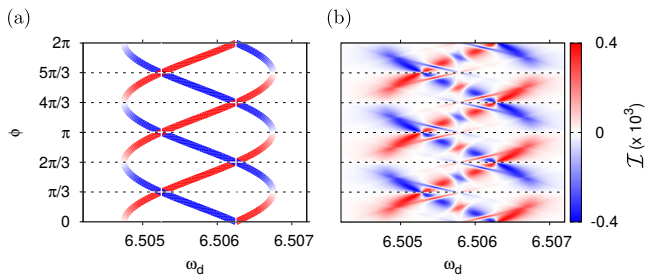}
 \caption{Nonequilibrium steady-state current $\mathcal{I}$ as a function of the microwave drive frequency $\omega_\rmd$ and the phase of the time-dependent couplers $\phi$ (energies in units of $2\pi$~GHz).
(a) Analytical prediction based on Eqs.~(\ref{eq:wopt}) and (\ref{eq:current}).
The horizontal dashed lines correspond to phases at which there cannot be a current.
(b) Numerical results from a Master Equation approach (see the text).
The main differences come from the population of higher-excited states carrying a non-vanishing current, which were neglected in the analytics.}
 \label{fig:current}
\end{center}
\end{figure}

\begin{figure}
\begin{center}
 \includegraphics[height=0.30\hsize, angle=-0]{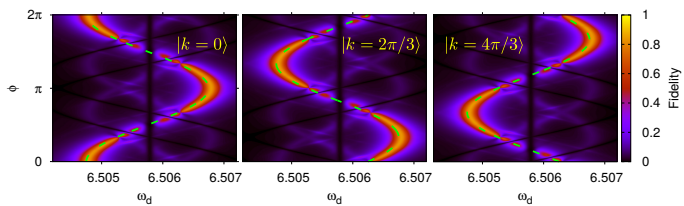}
 \caption{Steady-state population $n_k$ of the states  $|k=0 \rangle$,  $|k=2 \pi/3 \rangle$, and $|k=4 \pi/3 \rangle$ as a function of the microwave drive frequency $\omega_\rmd$ (in units of $2\pi$~GHz) and the phase of the time-dependent couplers $\phi$. The green dashed lines are the theoretical predictions for the optimal driving frequency $\omega_\rmd^{\rm opt}(\phi)$ in Eq.~(\ref{eq:wopt}).
%f A key element to the generation of a persistent current is to achieve current-carrying states with high fidelities.
}
 \label{fig:fidelities}
\end{center}
\end{figure}

 Using the framework developed above, we compute the non-equilibrium steady-state current of qubit excitations circulating around the ring, \textit{i.e}, we compute $\mathcal{I} = \mathrm{Tr}[I_i \rho_\sigma^{\infty}]$ where $I_i = -  \rmi J_0  \left[ \rme^{\rmi \phi} \sigma^+_{i} \sigma^-_{i+1}  - \mbox{h.c.} \right]$.
The current can be expressed in terms of the steady state populations (or fidelities) of the qubit eigenstates, $n_k(\omega_\rmd, \phi)  \equiv \langle k | \rho_\sigma^{\infty}  |  k  \rangle$, 
\begin{align} \label{eq:current}
\mathcal{I}(\omega_\rmd, \phi) =  \frac{2}{N} J_0  \sum_k \sin(k+\phi) n_k(\omega_\rmd, \phi)\,.
\end{align}
The above formula makes it transparent that a sizable current can be achieved by populating a specific qubit eigenstate $|k\rangle$ with high fidelity, $n_k \approx 1$, or by populating several neighboring $|k\rangle$ states close enough to contribute constructively to the overall current.  The latter regime is relevant for large rings, where the state-crowding in the one-excitation manifold prevents targeting a given state with high fidelity, but where a sizable current can be obtained as long as $\kappa \ll J_0$.
Equation~(\ref{eq:current}) also predicts a  magnitude for the permanent current  which is on the order of  $2 J_0/N$. This matches the values of the evanescent currents experimentally realized in Ref.~\cite{roushan}.

Let us now work out how the permanent current $\mathcal{I}$ behaves as a function of our two drive parameters: the microwave source frequency $\omega_\rmd$ and the phase of the time-dependent couplers $\phi$.
Let us first remark that at the specific values $\phi = n \pi/N$ with $n = 0 \ldots 2N-1$, the complex hopping amplitudes in $H_\sigma$ can be be made real by simple local unitary rotations, resulting in a time-reversal invariant qubit Hamiltonian which, therefore, cannot carry any current. In the case of $N=3$, this yields a vanishing current $\mathcal{I}(\omega_\rmd, \phi) = 0$ at $\phi = 0, \pi/3, 2\pi/3, 4\pi/3$ for any $\omega_\rmd$.
Away from those zero-current lines, the current is extremized
when the energy matching condition necessary for the Raman inelastic scattering processes is satisfied, \textit{i.e.} when the driving frequency is set to $\omega_\rmd^{\rm opt}(\phi)$ in Eq.~(\ref{eq:wopt}). There are $N$ of these curves, one for each value of $k$. Along those curves, assuming the state $|k\rangle$ is achieved with a near perfect fidelity, $n_k \approx 1$, the total current is given by $ \mathcal{I}(\omega_\rmd^{\rm opt}(\phi), \phi) \approx  \frac{2}{N} J_0  \sin(k+\phi)$. The energy matching condition above Eq.~(\ref{eq:wopt}) is expected to be valid within a frequency range set by the cavity decay loss, $\kappa$. Away from this, the qubits are in their ground state $| \boldsymbol{0} \rangle$ which does not carry any current.

 \paragraph*{Numerics.} 
We numerically control the different analytic approximations and assumptions made above (truncation of the Hilbert space, perturbation theory in the drive amplitude, perfect fidelities). We compute the current $\mathcal{I} = \mathrm{Tr}_\sigma [\mathcal{J}_i \rho_\sigma^{\infty}]$ by (i) performing an exact numerical diagonalization of the untruncated Hamiltonian $H_{\sigma}$ in Eq.~(\ref{eq:Hspinchain}), 
(ii) computing all the Fermi Golden Rule rates between the  $2^N$ qubit eigenstates, 
(iii) computing the steady-state density matrix by solving the Master Equation generalizing the one in Eq.~(\ref{eq:master}) to the full qubit Hilbert space.

The analytical and the numerical results for the current  $\mathcal{I}(\omega_\rmd, \phi)$ are gathered in Fig.~\ref{fig:current}. Figure~\ref{fig:fidelities} presents the steady-state populations $n_k(\omega_\rmd, \phi)$. 
By and large, the numerics confirm the analytic predictions, validating \textit{a posteriori} the different approximations and assumptions that were made. 
Figure~\ref{fig:current} also reveals new regions where the current is reversed compared to the predictions. These correspond to regimes where a resonance of the energy splittings between ground-state and single-excited states on the one hand, and between single-excited states and double-excited states on the other hand, is responsible for a nonvanishing overlap of the state of the system with the eigenstates of the two-excitation manifold.

\paragraph*{Conclusions and Outlook:}

We have successfully generated permanent spin currents in a superconducting-qubit architecture subject to dissipation. Our scheme is experimentally realizable and scalable. 
The drive frequencies $\omega_i^\rmd$ can be individually tuned dynamically to alleviate imperfections that may occur in the qubit or cavity frequencies. This relative robustness against imperfections is paramount for the study of topological phases and more generally for quantum computation with mesoscopic systems.
%  We have demonstrated that the combined effect carefully engineered drives and inevitable dissipation can lead to topological protection. 
A future interesting challenge is the enhancement of the current, further than the mega-excitation per second scale. This may be possible by using higher-excited chiral states, which carry more than one excitation around the ring. One way is to use ``colored'' cavity microwave drives with multiple frequencies finely tuned to steer the system to converge and remain in those higher-excited states~\cite{cami2}.

\textit{Acknowledgements: } CA and MK gratefully acknowledge the hospitality of the International Centre for Theoretical Sciences of Tata Institute of Fundamental Research, Bangalore, during the Open Quantum Systems program (ICTS/Prog-oqs/2017/07) where most of this research was completed.  MK gratefully acknowledges the Ramanujan Fellowship SB/S2/RJN-114/2016 from the Science and Engineering Research Board (SERB), Department of Science and Technology, Government of India. EK was supported by the Louisiana Board of Regents RCS grant LEQSF(2016-19)-RD-A-19, and by the National Science Foundation grant PHY-1653820.

\bibliography{compJ,fullbib}

\end{document}